\documentclass[aps,twocolumn,prl]{revtex4}
\usepackage{graphicx}
\usepackage{latexsym}
\usepackage{amsmath}
\usepackage{amsfonts}
\usepackage{amssymb}

\newcommand{\be}{\begin{equation}}
\newcommand{\ee}{\end{equation}}
\newcommand{\bea}{\begin{eqnarray}}
\newcommand{\eea}{\end{eqnarray}}
\begin{document}
\title{Metastable tight knots in a worm-like polymer }
\author{Alexander Y. Grosberg$^{\dag}$ and Yitzhak Rabin$^{\ddag}$ }
\affiliation{$^{\dag}$Department of Physics, University of Minnesota, 116 Church Street SE,
Minneapolis, MN 55455, USA}
\affiliation{$^{\ddag}$Department of Physics, Bar--Ilan University, Ramat-Gan 52900, Israel }
\date{\today}

\begin{abstract}
Based on an estimate of the knot entropy of a worm-like chain we predict that
the interplay of bending energy and confinement entropy will result in a
compact metastable configuration of the knot that will diffuse, without
spreading, along the contour of the semi-flexible polymer until it reaches one
of the chain ends. Our estimate of the size of the knot as a function of its
topological invariant (ideal aspect ratio) agrees with recent experimental
results of knotted dsDNA. Further experimental tests of our ideas are proposed.

\end{abstract}
\maketitle

While everyday experience suggests that knots are very common in long linear
strings of any kind, and they get quite tight whenever a string is not
carefully handled, the study of knots in polymers concentrated almost
exclusively on closed loops. It is understandable in the sense that knots are
not mathematically well defined for open strings. However, if the string is
long enough, while the knot occupies a short fragment of it far from the ends,
then distinguishing between different knots, or between knots and no knots,
becomes sufficiently unambiguous.

The recent achievement in the theory of knotted loops
\cite{Orlandini1, Metzler_Kardar_Knot_Localization, Orlandini2} is
the idea of knot localization. As simulations show
\cite{Orlandini2}, when a polymer loop with a knot is placed in a
good or $\theta$-solvent, it typically adopts a conformation in
which most of the polymer forms a long unknotted loop, while the
knot gets somewhat tightened in a small part of the contour. A
theoretical explanation of the knot localization phenomenon is given
\cite{Metzler_Kardar_Knot_Localization} for the so-called flat
knots, confined in a thin slit between two planes. In this case, one
can consider a two-dimensional network which corresponds to any
configuration of the knot by identifying chain crossings as the
cross-links. This mapping allows one to resort to the sophisticated
theory of 2D networks with excluded volume
\cite{Duplantier_2D_network,Duplantier_2D_network_details} and
conclude that the most entropically favorable network is obtained
when one of the knot arcs is made long at the expense of all others,
thus localizing the knot.

In the present note, we would like to use much simpler hand-waving
arguments to show that, at least for a worm-like polymer, there
exists a local (metastable) minimum of free energy corresponding to
a tight state of the knot even when the chain itself is open (not a
loop). That means that if we intentionally tie a sufficiently tight
knot somewhere on a very long polymer chain, the knot will
spontaneously shrink or expand to a well-defined size and then, on
much longer time scales, it will diffuse along the polymer (by
polymer self-reptating through the knot) until, finally, the knot is
released through the chain end. We expect that this knot will
diffuse along the polymer as a soliton, in the sense that its size
will remain relatively stable as it diffuses over large distances.
The size of such a solitary knot depends on the complexity of the
knot and on the persistence length of the polymer; we estimate that
the knot will tighten to a size smaller than the persistence length
of the polymer.

\begin{figure}
\centerline{\scalebox{0.4}{\includegraphics{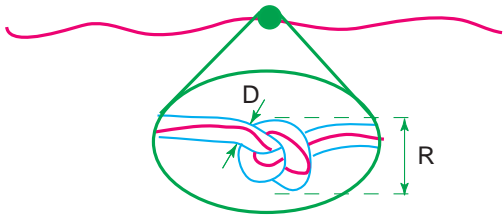}}}\caption{(Color
online) A knot is tied on a polymer chain. Knot size in space is
$R$, chain within the knot is confined to a self-consistent tube of
diameter $D$. Length of this tube is $pD$, where $p$ is the
topological invariant of the knot.} \label{fig:cartoon}
\end{figure}

To put forward our argument, we employ the knot entropy estimates
based on the idea \cite{Flory_Knot}, similar to that in reptation
theory \cite{Doi_Edwards} , that non-crossing constraints imposed on
the chain in the knot can be self-consistently described by
confining the chain in an effective tube. Physically most natural
construction of such a tube corresponds to making maximally inflated
or, equivalently, shortest length tube consistent with the given
knot topology. This approach to estimate knot entropy was invented
by us \cite{Flory_Knot}. It was also used in a number of other
contexts and widely popularized under the name ``ideal knots''
\cite{Ideal_Knots}. The configuration and aspect ratio, $p$, of the
``ideal'' tube represent topological invariants of the knot. This is
sketched in the figure \ref{fig:cartoon}.

Within the framework of this approach, we imagine that a tight knot,
characterized by ideal aspect ratio $p$, has been tied in a very long polymer,
and that currently the degree of tightening of the knot is such that the
diameter of its self-consistently confining tube is $D$. The length of the
tube is $pD$ and the size of the knot in space is $R$, such that $pD\times
D^{2}\sim R^{3}$ or
\begin{equation}
R\sim p^{1/3}D\ .\label{eq:densely_packed_tube}
\end{equation}
Throughout this paper, we drop all numerical coefficients, emphasizing only
the scaling aspects of our considerations. We want to estimate the free energy
of the knot as a function of $D$ or $R$.

Although the original estimates of knot entropy were designed for a
Gaussian chain \cite{Flory_Knot}, they can be readily adapted to a
worm-like polymer. 
Consider a knot with $D<\ell$, where $\ell$ is the persistence
length of the polymer. In this case, the polymer is quite tightly
confined in the tube and can only wiggle a little bit around its
centerline. Therefore, the length of polymer within the knot is
close to $pD$, and its free energy consists of bending energy and
confinement entropy. We estimate the bending energy as
$TpD\ell/R^{2}$, assuming temperature $T$ is expressed in energy
units ($k_B=1$), and assuming that the radius of curvature of the
tube is about the knot size $R$, which is natural for an ``ideal''
knot. We further assume that the confinement entropy is the same as
that for a straight tube for which case it was computed by Odijk
\cite{Odijk_deflection_length} and turns out to be about unity per
every so-called deflection length $\lambda\sim(D^{2}\ell)^{1/3}$;
this results in the free energy contribution about $ TpD/\lambda\sim
 Tp(D/\ell)^{1/3}$. Thus, using also
(\ref{eq:densely_packed_tube}), we obtain the following free energy
estimate:
\begin{equation}
\frac{\Delta F}{ T}\sim p^{1/3}\frac{\ell}{D}+p\left( \frac{D}{\ell
}\right)  ^{1/3}\ .\label{eq:free_energy}
\end{equation}
where $\Delta F$ is the free energy penalty for forming the knot (taking the
reference free energy to be that of the unknotted chain). Obviously, it has a
minimum at
\begin{equation}
D^{\ast}\sim\ell p^{-1/2}\ ,\ \ \ \ \mathrm{or}\ \ \
R^{\ast}\sim\ell p^{-1/6}\ .\label{eq:optimal}
\end{equation}
Notice that the resulting optimal $D^{\ast}$ meets the condition $D^{\ast
}<\ell$, so our estimate is self-consistent in this respect.

Of course, our result applies only as long as $D^{\ast}>d$, where
$d$ is the thickness of the polymer itself. Furthermore, in the
practically important case of a charged polymer, such as DNA, we
should also require that tube diameter exceeds Debye screening
length, $D^{\ast}>r_{s}$. In general, we can write roughly
$p<(\ell/(d+r_{s} ))^{2}$. More complex knots get so tight that
their further collapse is stopped by either the excluded volume or
electrostatic repulsion. As an example, for the dsDNA, the ratio
$\ell/(d+r_{s})\approx 20$ under physiological conditions
\cite{DNA_effective_diameter}, while all knots with 7 or fewer
crossings on the projection have $p$ about $30$ or less
\cite{Pieranski}. Therefore, in practice the condition on knot
complexity $p<(\ell/(d+r_{s} ))^{2}$ is not very restrictive.

Let us now try to understand the physical meaning of our result
(\ref{eq:optimal}), because at the first glance it might seem
counterintuitive. Indeed, the optimal $D^{\ast}$ results from the
competition of two factors, each of which, as it seems, disfavors
tightening! One, chain bending energy, obviously favors more loose
states of the knot, or increasing $D$. The other one, however,
related to the confinement entropy, favors tube widening \textit{if}
the chain length in the tube is fixed. In our case, when a knot
tightens, it reduces the length of the chain, $pD$, confined in the
knot. In other words, the part of the chain that remains in the tube
suffers more when $D$ decreases, but the other part of the chain
gets completely free of restriction, and that factor wins the whole
game. That means, what really tightens the knot is the entropy gain
of the chain tails outside of the knot. Alternatively, we can say
that the knot gets compressed by the pressure of Rouse modes (or
bending phonons) of the outside chain tails. Similar force can be
observed during the translocation of a long polymer through a narrow
channel across a membrane; while two long polymer ends are outside
the membrane, their entropic favorability results in a force
stretching the polymer portion inside the channel, and, accordingly,
compressing the membrane. Notice that while in this planar geometry
this force increases logarithmically with the length of the polymer
tails, in our case of two long polymer ends sticking out of a
compact knot, the force on the knot is independent of the length of
these ends.

How tight should be the knot in the first place in order for our
mechanism to take over and to bring the knot to its metastable size
$R^{\ast}$? In other words, how wide is the basin of attraction of
our metastable free energy minimum? We argue that the knot should be
initially tightened to the state in which tube diameter is smaller
or about chain persistence length $\ell$. Indeed, when knot tube is
wider, $D>\ell$, the chain inside the tube is roughly Gaussian, with
blobs of size $D$. Each blob contains about $D^{2} /\ell$ of polymer
contour length, and the entire tube contains polymer length
$pD^{2}/\ell$. At the same time, confinement entropy is about unity
per blob, which results in overall confinement entropy of about $p$,
independent of $D$. Thus, our entropic knot tightening effect does
not work if the tube is wider than persistence length, and it comes
into play only when the knot is prepared in a compact enough state
such that $D<\ell$. To achieve this, the knot should be initially
prepared by the pulling the ends with force $f>f^{\ast}$, where
$f^{\ast}\sim  T/\ell$.

Let us discuss now the dynamics aspect of the situation. Imagine
once again that a sufficiently tight knot was initially tied in the
polymer, similar to how it was done in the experimental work
\cite{Quake_Knot_experiment} (see also
\cite{Quake_Knot_experiment_Erratum}) with DNA. Suppose now that the
chain is released and is free to move. We predict then that the knot
tightens spontaneously within a time which is roughly independent of
the total chain length $L$. After that, the knot will diffuse along
the chain in pretty much the same way as it was observed
experimentally
\cite{Quake_Knot_experiment,Quake_Knot_experiment_Erratum} and
numerically \cite{Vologodski_knot_diffusion} for the stretched chain
(even though in the experiment of ref. \cite{Quake_Knot_experiment}
the chain ends were held at fixed separation throughout the
diffusion process). As regards the knot diffusion coefficient along
the chain, it was shown in the work \cite{Quake_Knot_experiment}
that it can be quite accurately expressed in terms of the friction
coefficient, $\zeta$, of a polymer with length $pD$ and diameter
$d$, moving in the viscous solvent inside a tube of diameter $D$:
$\zeta=\frac{2\pi\eta}{\ln(D/d)}pD$. 
The diffusion coefficient is then determined by Einstein's relation
as $ T/\zeta$, and the full relaxation time of knot diffusion to the
chain end is about $L^{2}\zeta/ T$. In these formulae, $\eta$ is
assumed to be the bulk viscosity of the solvent.

\begin{figure}
\centerline{\scalebox{0.4}{\includegraphics{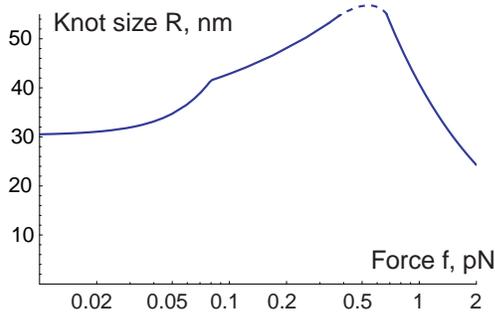}}}\caption{(Color
online) Knot size $R^{\ast}$ against the stretching force $f$. Force
is indicated in piconewtons and knot size in nanometers, both under
the assumptions that $\ell= 50 \ nm$, and that there are no any
numerical coefficients to worry about.} \label{fig:R_against_force}
\end{figure}

The above description of a soliton-like knot diffusing as a whole
along the chain holds as long as diffusion time is shorter than the
time of thermally-activated loosening of the knot. Indeed, we argued
that the knot gets to its metastable size $R^{\ast}$ only when it is
compact enough to begin with, such that $D < \ell$ and, therefore, a
free energy barrier exists at $D \approx \ell$ (knots with $D>\ell$
will loosen up spontaneously). Using formula (\ref{eq:free_energy}),
we can estimate the barrier height as $\left. \Delta F\right\vert
_{D/\ell=1}-\left.  \Delta F\right\vert _{D=D^{\ast}}\sim
 T\left(  p-p^{5/6}\right)  $. As one could have expected, this
barrier is very high for complex knots, but even for $p \approx 30$,
one gets a barrier of about $10  T$ which should be sufficient to
keep a knot locked while diffusing over a distance of few hundred
persistence lengths, typical of DNA manipulation experiments.
For smaller values of $p$, the barrier may be too small to stabilize
the tight knot and without the applied stretching it will spread due
to thermal fluctuations. 

As a corollary to our result, let us mention the following rather unexpected
prediction. Let us take the polymer with the knot tightened according to our
mechanism, and now let us gently pull the chain ends by a weak force. The
applied stretching will suppress transverse fluctuations of the open chain
ends. But these fluctuations are exactly the reason why the knot was tightened
in the first place. Therefore, when these fluctuations are suppressed by the
applied force, the knot swells - instead of further tightening which one could
have naively expected. Of course, at the larger forces the loosening of the
knot stops and normal tightening takes over.

Let us support this physical argument by a little calculation. If
the chain of contour length $L$ is stretched by a weak force $f$, it
represents the succession of Pincus blobs, each involving the number
of persistence lengths $g$ such that $f\ell g^{1/2}\sim  T$; the
stretching free energy is about $ T$ per blob, or $ TL/g\ell\sim
L\ell f^{2}/ T$ (see, e.g., \cite{deGennes_book} for further
details). In our case, only the chain outside of the knot is subject
to this stretching effect, which yields the overall free energy
\begin{equation}
\frac{\Delta F}{ T}\sim p^{1/3}\frac{\ell}{D}+p\left(  \frac{D}{\ell
}\right)  ^{1/3}+\frac{L-pD}{\ell}\left(  \frac{\ell f}{ T}\right)
^{2}\ , \label{eq:to_minimize_Pincus_regime}
\end{equation}
subject to optimization with respect to $D$ (see Eq.
(\ref{eq:result_of_optimization}) below); here, the term $ \propto
L$ is large, but can be dropped as independent of $D$. Formula
(\ref{eq:to_minimize_Pincus_regime}) remains valid as long as Pincus
blob is larger than persistence length $f\ell/ T<1$.

If we stretch the outside chain even further, beyond the Gaussian
regime, it crosses-over to the so-called Marco-Siggia or Odijk
regime in which stretching free energy is about $ T$ per Odijk
deflection length. In this regime,
\begin{equation}
\frac{\Delta F}{ T}\sim p^{1/3}\frac{\ell}{D}+p\left( \frac{D}{\ell
}\right)  ^{1/3}+\frac{L-pD}{\ell}\left(  \frac{\ell f}{ T}\right)
^{1/2}\ ; \label{eq:to_minimize_Odijk_regime}
\end{equation}
once again, this has to be optimized with respect to $D$.

Notice that in both equations (\ref{eq:to_minimize_Pincus_regime}) and
(\ref{eq:to_minimize_Odijk_regime}) we have neglected the effect of force on
the chain part inside of the knot. This is justified as long as the amount of
transverse fluctuations of the outside chain, $D_{\mathrm{out}}(f)$, which is
either the blob size in Pincus regime or the tube diameter in Odijk regime,
remains smaller than the optimized tube diameter inside the knot, $D^{\ast}$.
At still larger forces, the chain inside the knot is just as stretched as
outside. Therefore, we can summarize all results of optimization in the
following way:
\begin{equation}
R^{\ast}\sim\left\{
\begin{array}
[c]{lcr} \ell p^{-1/6}+\ell p^{-1/2}\left( \frac{f\ell}{ T}\right)
^{2} &
\mathrm{for} & \frac{f\ell}{ T}<1\\
\ell p^{-1/6}+\ell p^{-1/2}\left(  \frac{f\ell}{ T}\right)  ^{1/2} &
\mathrm{for} & 1<\frac{f\ell}{ T}<p^{1/3}\\
\ell p^{1/3}\left(  \frac{f\ell}{ T}\right)  ^{-3/4} & \mathrm{for}
& p^{1/3}<\frac{f\ell}{ T}
\end{array}
\right.  \label{eq:result_of_optimization}
\end{equation}
These results are sketched in figure \ref{fig:R_against_force}. In accordance
with our qualitative argument, the application of weak force loosens the tight
knot instead of tightening it further!

In practice, for the case of dsDNA, since $\ell\approx50 \ nm$ and $
T \approx4 \ pN \times nm$, and taking $p \approx30$, we get that
both cross-overs $f \ell/ T \sim1$ and $f \ell/ T \sim p^{1/3}$ are
within an experimentally feasible force range (about a few tenth and
about a few piconewtons, respectively).

\begin{figure}
\centerline{\scalebox{0.4}{\includegraphics{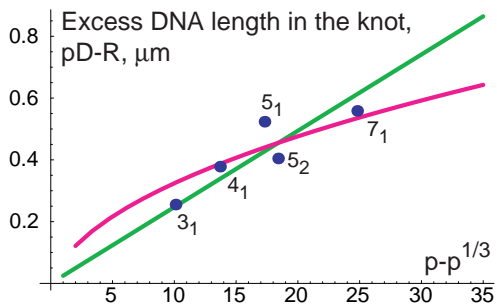}}}\caption{(Color
online) The excess DNA length in the knot, $p D^{\ast} - R^{\ast}$
against $p-p^{1/3}$. The meaning of these variables is explained in
the text. Data points are taken from the work
\cite{Quake_Knot_experiment,Quake_Knot_experiment_Erratum}. Linear
fit corresponds to the assumption that $D^{\ast}$ does not depend on
the knot type, on $p$. Curved line corresponds to our theory.}
\label{fig:D_against_p}
\end{figure}

Speaking about experimental tests of our theory, we should of course
consider first the experiment by Quake et al
\cite{Quake_Knot_experiment,Quake_Knot_experiment_Erratum} in which
the knots were tied on DNA by the use of optical tweezers and then
the knot motion along DNA was observed. Our theory indicates that
the optimal tube diameter in the knot, $D^{\ast}$, does depend in a
certain way on the knot complexity, on $p$. In the experiment,
authors were unable to observe directly the knot and measure its
size $R$, but they were able to detect the amount of fluorescence
coming from the knot region in excess of the fluorescence from
linear DNA. That means, in our notations, that authors measured
quantity $pD^{\ast}-R^{\ast}$, because $pD^{\ast}$ is the length of
DNA within the knot (proportional to the amount of fluorescence),
while $R$ is the length of DNA that would have been in that place if
there were no knot. Given the normalization condition
(\ref{eq:densely_packed_tube}), one could write
$pD^{\ast}-R^{\ast}=\left(  p-p^{1/3}\right)  D^{\ast}$. Authors of
the work \cite{Quake_Knot_experiment}, assuming \textit{a'priori}
that $D$ is independent of $p$, plotted $pD^{\ast}-R^{\ast}$ against
$p-p^{1/3}$, fitted the data to the linear function, and interpreted
its slope as the tube diameter $D^{\ast}$. By contrast, our theory
predicts that $D^{\ast}$ does depend on knot complexity $p$, such
that $pD^{\ast}-R^{\ast}=\left( p^{1/2}-p^{-1/6}\right) \ell$. This
is illustrated in the figure \ref{fig:D_against_p}, where both the
linear fit and our theory results are plotted along with the data
points reproduced from the work \cite{Quake_Knot_experiment}. At
present, it is impossible to decide which theory is a better fit to
the data.

To further test our predictions, one should prepare the knot and
then consider what happens to it if the force is subsequently
switched off or significantly reduced. Another test of our theory
would be to see how the knot diffusion time, determined by 
the friction coefficient $\zeta$, depends on the knot type through
$D$. Yet another possibility is to use granular macroscopic chain
experiments, as described in \cite{Ben-Naim_Granular_Chain}. In the
movie available on the site
\begin{verbatim}http://cnls.lanl.gov/~ebn/research/pics/chain.mov\end{verbatim}
the knot does move along the chain like a soliton, it does not
increase in size while moving, in accord with our theory. It would
be interesting to make a detailed statistics on this subject,
although we should also put in question the applicability of our
arguments based on worm-like chain model to the granular chain
examined in \cite{Ben-Naim_Granular_Chain}.

To conclude, we have shown that a sufficiently complex knot in a
worm-like polymer will attain a well-defined compact size which is
smaller that the persistence length.  This configuration is
metastable and the knot will diffuse along the chain as a soliton,
whether the ends of the chain are subjected to external force or are
free to move. Notice that since the results depend crucially on the
elastic properties of the polymer on length scales comparable to and
smaller than the persistence length, the effects described in this
work are quite sensitive to the choice of the polymer model and can
not be captured by locally stiff  (e.g., freely jointed or freely
rotating) chain models. Likewise, solitary knots will not form in
Gaussian chains for which the persistence length $\ell$ and the
thickness of the chain $d$ coincide. Such knots are expected to form
in semi-flexible polymers such as double stranded DNA for which (a)
separation of length scales exists, $L \gg \ell \gg d$ and (b)
elastic behavior on length scales smaller than $\ell$ was
demonstrated in single molecule experiments \cite{Bustamante,
Bensimon}.

\acknowledgements We would like to acknowledge useful discussion
with Eli Ben-Naim. We thank Xiaoyan Robert Bao for providing us with
the data points used above in Figure \ref{fig:D_against_p}. The work
of AG was supported in part by the MRSEC Program of the National
Science Foundation under Award Number DMR-0212302. This research was
supported in part by a grant from the US-Israel Binational Science
Foundation.

\bibliography{TightKnots}

\end{document}